\documentclass[aps,pre,twocolumn,superscriptaddress,showpacs,showkeys]{revtex4}

\usepackage{natbib}
\usepackage{dcolumn}
\usepackage{amsmath}
\usepackage{graphicx}
\usepackage{rotating}
\usepackage{multirow}

\begin{document}

\title{Resolution limit in community detection}

\author{Santo Fortunato}
\affiliation{School of Informatics and Biocomplexity Center,
Indiana University,Eigenmann Hall, 1900 East Tenth Street,
Bloomington IN 47406}
\affiliation{Fakult\"at f\"ur Physik, Universit\"at Bielefeld, D-33501 Bielefeld, Germany}
\affiliation{Complex Networks Lagrange Laboratory (CNLL), ISI Foundation, Torino, Italy}

\author{Marc Barth\'elemy}
\affiliation{School of Informatics and Biocomplexity Center,
Indiana University,Eigenmann Hall, 1900 East Tenth Street,
Bloomington IN 47406}
\affiliation{CEA-Centre d'Etudes de
Bruy{\`e}res-le-Ch{\^a}tel, D\'epartement de Physique Th\'eorique et
Appliqu\'ee BP12, 91680 Bruy\`eres-Le-Ch\^atel, France}

\date{\today} \widetext

\begin{abstract}

  Detecting community structure is fundamental to clarify the link
  between structure and function in complex networks and is used for
  practical applications in many disciplines.  A successful method
  relies on the optimization of a quantity called modularity [Newman
  and Girvan, Phys. Rev. E {\bf 69}, 026113 (2004)], which is a
  quality index of a partition of a network into communities. We find
  that modularity optimization may fail to identify modules smaller
  than a scale which depends on the total number $L$ of links of the
  network and on the degree of interconnectedness of the modules, even in cases
  where modules are unambiguously defined. The probability that a module
  conceals well-defined substructures is the highest if the number of links
  internal to the module is of the order of $\sqrt{2L}$ or smaller.  We discuss the
  practical consequences of this result by analyzing partitions
  obtained through modularity optimization in artificial and real
  networks.

\end{abstract}

\pacs{89.75.-k, 89.75.Hc, 05.40 -a, 89.75.Fb, 87.23.Ge}
\keywords{Networks, community structure, modularity}
\maketitle



\section{Introduction}

Community detection in complex networks has attracted a lot of
attention in the last years (for a review
see~\cite{Newman:2004,Danon:2005}). The main reason is that complex
networks~\cite{bara02,mendes03,Newman:2003,psvbook,vitorep} are made
of a large number of nodes and that so far most of the quantitative
investigations were focusing on statistical properties disregarding
the roles played by specific subgraphs. Detecting communities (or
\textit{modules}) can then be a way to identify relevant substructures
that may also correspond to important functions. In the case of the
World Wide Web, for instance, communities are sets of Web pages
dealing with the same topic~\cite{Flake:2002}. Relevant community
structures were also found in social networks~\cite{Girvan:2002,
  Lusseau:2003, Adamic:2005}, biochemical
networks~\cite{Holme:2003,Guimera:2005,palla}, the
Internet~\cite{Eriksen:2003}, food webs~\cite{foodw}, and in networks
of sexual contacts~\cite{sexcontact}.

Loosely speaking a community is a subgraph of a network whose nodes
are more tightly connected with each other than with nodes outside the
subgraph. A decisive advance in community detection was made by Newman
and Girvan~\cite{Newman:2004b}, who introduced a quantitative measure
for the quality of a partition of a network into communities, the
so-called \textit{modularity}. This measure essentially compares the
number of links inside a given module with the expected value for a
randomized graph of the same size and degree sequence.  If one takes
modularity as the relevant quality function, the problem of community
detection becomes equivalent to modularity optimization. The latter is
not trivial, as the number of possible partitions of a network in
clusters increases exponentially with the size of the network, making
exhaustive optimization computationally unreachable even for
relatively small graphs. Therefore, a number of algorithms have been
devised in order to find a good optimization with the least
computational cost. The fastest available procedures uses greedy
techniques~\cite{Newman:2004c, Clauset:2004} and extremal
optimization~\cite{Duch:2005}, and are at present time the only
algorithms capable to detect communities on large networks.  More
accurate results are obtained through simulated
annealing~\cite{Guimera:2004,Reichardt:2006}, although this method is
computationally very expensive.

Modularity optimization seems thus to be a very effective method to
detect communities, both in real and in artificially generated
networks. The modularity itself has however not yet been thoroughly
investigated and only a few general properties are known. For example,
it is known that the modularity value of a partition does not have a
meaning by itself, but only if compared with the corresponding
modularity expected for a random graph of the same
size~\cite{Bornholdt:2006}, as the latter may attain very high values,
due to fluctuations~\cite{Guimera:2004}.

In this paper we focus on communities defined by modularity. We
will show that modularity contains an intrinsic scale which depends on
the number of links of the network, and that modules smaller than that
scale may not be resolved, even if they were complete graphs connected
by single bridges. The resolution limit of modularity actually depends
on the degree of interconnectedness between pairs of communities and can reach
values of the order of the size of the whole network. It is thus
\textit{a priori} impossible to tell whether a module (large or
small), obtained through modularity optimization, is indeed a single
module or a cluster of smaller modules. This result thus introduces
some caveats in the use of modularity to detect community structure.

In Section~\ref{sec2} we recall the notion of modularity and discuss
some of its properties. Section~\ref{sec3} deals with the problem of
finding the most modular network with a given number of nodes and
links. In Section~\ref{sec4} we show how the resolution limit of
modularity arises. In Section~\ref{sec5} we illustrate the problem
with some artificially generated networks, and extend the discussion
to real networks. Our conclusions are presented in Section~\ref{sec6}.

\section{Modularity}
\label{sec2}

The modularity of a partition of a network in modules can be written
as~\cite{Newman:2004b}
\begin{equation}
Q=\sum_{s=1}^{m}\Big[\frac{l_s}{L}-\left(\frac{d_s}{2L}\right)^2\Big],
\label{eq:mod}
\end{equation}
where the sum is over the $m$ modules of the partition, $l_s$ is the
number of links inside module $s$, $L$ is the total number of links in
the network, and $d_s$ is the total degree of the nodes in module
$s$. The first term of the summands in Eq.~(\ref{eq:mod}) is the
fraction of links inside module $s$; the second term instead
represents the expected fraction of links in that module if links were
located at random in the network (under the only constraint that the
degree sequence coincides with that in the original
graph). If for a subgraph ${\cal S}$ of a network the first term is
much larger than the second, it means that there are many more links
inside ${\cal S}$ than one would expect by random chance, so 
${\cal S}$ is indeed a module. The comparison with the null model
represented by the randomized network leads to the quantitative
definition of community embedded in the \textit{ansatz} of
Eq.~(\ref{eq:mod}). We conclude that, in a modularity-based framework,
a subgraph ${\cal S}$ with $l_s$ internal links and total degree $d_s$
is a module if
\begin{equation}
\frac{l_s}{L}-\left(\frac{d_s}{2L}\right)^2>0.
\label{eq2}
\end{equation}
Let us express the number of links $l_s^{out}$ joining nodes of the
module $s$ to the rest of the network in terms of $l_s$,
i.e. $l_s^{out}=al_s$ with $a\geq 0$. So,
$d_s=2l_s+l_s^{out}=(a+2)l_s$ and the condition (\ref{eq2}) becomes
\begin{equation}
\frac{l_s}{L}-\left[\frac{(a+2)l_s}{2L}\right]^2>0,
\label{eq2bis}
\end{equation}
from which, rearranging terms, one obtains
\begin{equation}
l_s<\frac{4L}{(a+2)^2}.
\label{eq2ter}
\end{equation}
If $a=0$, the subgraph ${\cal S}$ is a disconnected part of the
network and is a module if $l_s<L$ which is always true. If $a$ is
strictly positive, Eq.~(\ref{eq2ter}) sets an upper limit to the
number of internal links that ${\cal S}$ must have in order to be a
module. This is a little odd, because it means that the definition of
community implied by modularity depends on the size of the whole
network, instead of involving a ``local'' comparison between the
number of internal and external links of the module. For $a<2$ one
has $2l_s>l_s^{out}$, which means that the total degree internal to
the subgraph is larger than its external degree, i.e.
$d^{in}_s>d^{out}_s$. The attributes ``internal'' and ``external''
here mean that the degree is calculated considering only the internal
or the external links, respectively. In this case, the subgraph ${\cal S}$ 
would be a community according to the ``weak'' definition given
by Radicchi et al.~\cite{radicchi}.

For $a<2$ the right-hand-side of inequality~(\ref{eq2ter}) is 
in the interval $[L/4,L]$.
A subgraph of size $l_s$ would then be a community both
within the modularity framework and
according to the weak definition of Radicchi et al. if $a<2$ and
$l_s$ is less than a quantity in the interval $[L/4,L]$. Sufficient
conditions for which these constraints are always met are then
\begin{equation}
l_s<\frac{L}{4}, a<2.
\label{eq2quater}
\end{equation}
In Section~\ref{sec4} we shall only
consider modules of this kind.
 
According to Eq.~(\ref{eq2}), a partition of a network into actual
modules would have a positive modularity, as all summands in
Eq.~(\ref{eq:mod}) are positive.  On the other hand, for particular
partitions, one could bump into values of $Q$ which are negative.  The
network itself, meant as a partition with a single module, has
modularity zero: in this case, in fact, $l_1=L$, $d_1=2L$, and the
only two terms of the unique module in $Q$ cancel each other. Usually,
a value of $Q$ larger than $0.3-0.4$ is a clear indication that the
subgraphs of the corresponding partition are modules. However, the
maximal modularity differs from a network to another and depends
on the number of links of the network. In the next section we shall
derive the expression of the maximal possible value $Q_M(L)$ that $Q$
can attain on a network with $L$ links. We will prove that the upper limit
for the value of modularity for any network is $1$ and we will see why the modularity
is not scale independent.

\section{The most modular network}
\label{sec3}

In this section we discuss of the most modular network which will
introduce naturally the problem of scales in modularity optimization.
In Ref.~\cite{Danon:2005}, the authors consider the interesting
example of a network made of $m$ identical complete graphs (or
`cliques'), disjoint from each other. In this case, the modularity is
maximal for the partition of the network in the cliques and is given
by the sum of $m$ equal terms. In each clique there are $l=L/m$ links,
and the total degree is $d=2l$, as there are no links connecting nodes
of the clique to the other cliques. We thus obtain
\begin{equation}
Q=m\Big[\frac{l}{L}-\left(\frac{2l}{2L}\right)^2\Big]=
m\left(\frac{1}{m}-\frac{1}{m^2}\right)=1-\frac{1}{m},
\label{eq3}
\end{equation}
which converges to $1$ when the number of cliques goes to infinity.
We remark that for this result to hold it is not necessary that the
$m$ connected components be cliques. The number of nodes of the
network and  within the modules does not affect modularity. If we
have $m$ modules, we just need to have $L/m$ links inside the modules,
as long as this is compatible with topological constraints, like
connectedness. In this way, a network composed by
$m$ identical trees (in graph theory, a forest) has the same maximal
modularity reported in Eq.~(\ref{eq3}), although it has a far smaller
number of links as compared with the case of the densely connected
cliques (for a given number of nodes).

A further interesting question is how to design a \textit{connected}
network with $N$ nodes and $L$ links which maximizes modularity.
To address this issue, we proceed in two steps: first, we consider the
maximal value $Q_M(m,L)$ for a partition into a fixed number $m$ of
modules; after that, we look for the number $m^\star$ that maximizes
$Q_M(m,L)$.
  
\begin{figure}[t]
\includegraphics[width=5cm]{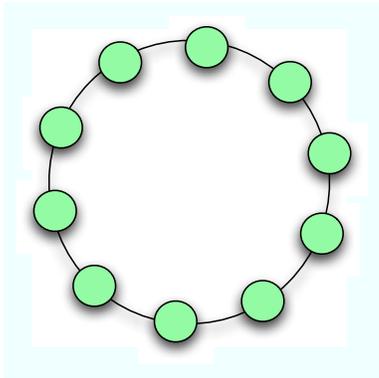}
\caption{\label{fig1} Design of a connected network with maximal modularity.
The modules (circles) must be connected to each other by the minimal number of links.}
 \end{figure}

Let us first consider a partition into $m$ modules. Ideally, to
maximize the contribution to modularity of each module, we should
reduce as much as possible the number of links connecting modules. If
we want to keep the network connected, the smallest number of
inter-community links must be $m-1$. For the sake of clarity, and to
simplify the mathematical expressions (without affecting the final
result), we assume instead that there are $m$ links between the
modules, so that we can arrange the latter in the simple ring-like
configuration illustrated in Fig.~\ref{fig1}.
The modularity of such a network is
\begin{equation}
Q=\sum_{s=1}^{m}\Big[\frac{l_s}{L}-\left(\frac{2l_s+2}{2L}\right)^2\Big],
\label{eq4}
\end{equation}
where 
\begin{equation}
\sum_{s=1}^{m}l_s=L-m.
\label{eq5}
\end{equation}
It is easy to see that the expression of Eq.~(\ref{eq4}) reaches its
maximum when all modules contain the same number of links, 
i.e. $l_s=l=L/m-1, \forall s=1,2,...,m$. The maximum is then given by
\begin{equation}
Q_M(m,L)=m\Big[\frac{L/m-1}{L}-\left(\frac{L/m}{L}\right)^2\Big]=1-\frac{m}{L}-\frac{1}{m}.
\label{eq6}
\end{equation}
We have now to find the maximum of $Q_M(m,L)$ when the number of
modules $m$ is variable. For this purpose we treat $m$ as a real
variable and take the derivative of $Q_M(m,L)$ with respect to $m$
\begin{equation}
\frac{dQ_M(m,L)}{dm}=-\frac{1}{L}+\frac{1}{m^2}
\label{eq7}
\end{equation}
which vanishes when $m=m^\star=\sqrt{L}$. This point indeed
corresponds to the absolute maximum $Q_M(L)$ of the function
$Q_M(m,L)$. This result coincides with the one found by
the authors of~\cite{Guimera:2004} for a one-dimensional lattice, 
but our proof is completely general and does not require preliminary
assumptions on the type of network and modules.

Since $m$ is not a real number, the actual maximum is reached when $m$
equals one of the two integers closest to $m^\star$, but that is not
important for our purpose, so from now on we shall stick to the
real-valued expressions, their meaning being clear. The maximal
modularity is then
\begin{equation}
Q_M(L)=Q_M(m^\star,L)=1-\frac{2}{\sqrt{L}},
\label{eq8}
\end{equation}
and approaches $1$ if the total number of links $L$ goes to infinity.
The corresponding number of links in each module is $l=\sqrt{L}-1$.
The fact that all modules have the same number of links does not imply
that they have as well the same number of nodes. Again, modularity
does not depend on the distribution of the nodes among the modules as
long as the topological constraints are satisfied. For instance, if we assume
that the modules are connected graphs, there must be at most
$n=l+1=\sqrt{L}$ nodes in each module. The crucial point here is that
modularity seems to have some intrinsic scale of order $\sqrt{L}$,
which constrains the number and the size of the modules. 
For a given total number of nodes and links we could build 
many more than $\sqrt{L}$ modules, but the corresponding network
would be less ``modular'', namely with a value of the
modularity lower than the maximum of Eq.~(\ref{eq8}). 
This fact is the basic reason why small modules may not be
resolved through modularity optimization, as it will be clear in the
next section.

\section{The resolution limit}
\label{sec4}

\begin{figure}[t]
\includegraphics[width=8cm]{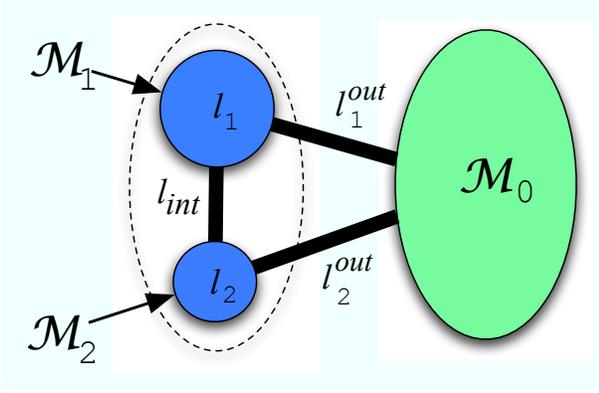}
\caption{\label{fig2} Scheme of a network partition into three or more
  modules. The two circles on the left picture two modules, the oval
  on the right reprensents the rest of the network ${\cal M}_0$, whose
  structure is arbitrary.}
\end{figure}

We analyze a network with $L$ links and with at least three modules,
in the sense of the definition of formula~(\ref{eq2quater})
(Fig.~\ref{fig2}).  We focus on a pair of modules, ${\cal M}_1$ and
${\cal M}_2$, and distinguish three types of links: those internal to each of
the two communities ($l_1$ and $l_2$, respectively), between ${\cal M}_1$ and
${\cal M}_2$ ($l_{int}$) and between the two communities 
and the rest of the network ${\cal M}_0$ ($l_1^{out}$ and
$l_2^{out}$). In order to simplify the calculations we express the numbers of
external links in terms of $l_1$ and $l_2$, so
$l_{int}=a_1l_1=a_2l_2$, $l_1^{out}=b_1l_1$ and $l_2^{out}=b_2l_2$,
with $a_1, a_2, b_1, b_2\geq 0$. Since ${\cal M}_1$ and ${\cal M}_2$ are modules
by construction, we also have $a_1+b_1\leq 2$, $a_2+b_2\leq 2$ and $l_1,l_2<L/4$
(see Section~\ref{sec2}).
Now we consider two partitions $A$ and $B$ of the network. In
partition $A$, ${\cal M}_1$ and ${\cal M}_2$ are taken as separate
modules, and in partition $B$ they are considered as a single
community. The split of the rest of the network is arbitrary but
identical in both partitions. We want to compare the modularity values
$Q_A$ and $Q_B$ of the two partitions.  Since the modularity is a sum
over the modules, the contribution of ${\cal M}_0$ is the same in both
partitions and is denoted by $Q_0$. From Eq.~(\ref{eq:mod}) we obtain
\begin{eqnarray}
Q_A&=&Q_0+\frac{l_1}{L}-\left[\frac{(a_1+b_1+2)l_1}{2L}\right]^2+\frac{l_2}{L}+\nonumber\\ 
& &-\left[\frac{(a_2+b_2+2)l_2}{2L}\right]^2;
\label{eq9}
\end{eqnarray}
\begin{eqnarray}
Q_B&=&Q_0+\frac{l_1+l_2+a_1l_1}{L}+\nonumber\\
&&-\left[\frac{(2a_1+b_1+2)l_1+(b_2+2)l_2}{2L}\right]^2.
\label{eq10}
\end{eqnarray}
The difference ${\Delta}Q=Q_B-Q_A$ is
\begin{equation}
{\Delta}Q=\left[2La_1l_1-
(a_1+b_1+2)(a_2+b_2+2)l_1l_2\right]/(2L^2).
\label{eq11}
\end{equation}
As ${\cal M}_1$ and ${\cal M}_2$ are both modules by construction, we
would expect that the modularity should be larger for the partition
where the two modules are separated, i.e. $Q_A>Q_B$, which in turn
implies ${\Delta}Q<0$. From Eq.~(\ref{eq11}) we see that ${\Delta}Q$
is negative if
\begin{equation}
l_2>\frac{2La_1}{(a_1+b_1+2)(a_2+b_2+2)}.
\label{eq12}
\end{equation}
We see that if $a_1=a_2=0$, i.e. if there are no links between 
${\cal M}_1$ and ${\cal M}_2$, the above condition is trivially
satisfied. Instead, if the two modules are connected to each other,
something interesting happens. Each of the coefficients $a_1$, $a_2$, $b_1$,
$b_2$ cannot exceed $2$ and $l_1$ and $l_2$ are both smaller than
$L/4$ by construction but can be taken as small as we wish with
respect to $L$. In this way, it is possible to choose $l_1$ and $l_2$
such that the inequality of Eq.~(\ref{eq12}) is not satisfied. In such
a situation we can have $\Delta Q>0$ and the modularity of the
configuration where the two modules are considered as a single
community is larger than the partition where the two modules are
clearly identified. This implies that by looking for the maximal modularity, there
is the risk to miss important structures at smaller scales.  To give
an idea of the size of $l_1$ and $l_2$ at which modularity
optimization could fail, we consider for simplicity the case in which
${\cal M}_1$ and ${\cal M}_2$ have the same number of links,
i.e. $l_1=l_2=l$. The condition on $l$ for the modularity to miss the
two modules also depends on the fuzziness of the modules, as expressed
by the values of the parameters $a_1$, $a_2$, $b_1$, $b_2$.  In order
to find the range of potentially ``dangerous'' values of $l$, we
consider the two extreme cases in which
\begin{itemize}
\item{the two modules have a perfect balance between internal and external degree
($a_1+b_1=2$, $a_2+b_2=2$), so they are on the edge between 
being or not being communities, in the weak sense defined in ~\cite{radicchi};}
\item{the two modules have the smallest possible external degree, which 
means that there is a single link connecting them to the rest of the network and 
only one link connecting each other ($a_1=a_2=b_1=b_2=1/l$).} 
\end{itemize}
In the first case, the maximum value that the coefficient of $L$ can take in
Eq.~(\ref{eq12}) is $1/4$, when $a_1=a_2=2$ and $b_1\approx 0$, $b_2\approx 0$, so
we obtain that Eq.~(\ref{eq12}) may not be satisfied for
\begin{equation}
l<l_R^{max}=\frac{L}{4},
\label{eq14}
\end{equation}
which is a scale of the order of the size of the whole network. In
this way, even a pair of large communities may not be resolved if they
share enough links with the nodes outside them (in this case we speak of 
``fuzzy'' communities). A more striking result emerges when we
consider the other limit, i.e. when $a_1=a_2=b_1=b_2=1/l$.  In this
case it is easy to check that Eq.~(\ref{eq12}) is not satisfied for
values of the number of links inside the modules satisfying
\begin{equation}
l<l_R^{min}=\sqrt{\frac{L}{2}}.
\label{eq15}
\end{equation}
If we now assume that we have two (interconnected) modules with the
same number of internal links $l<l_R^{min}<l_R^{max}$, the discussion
above implies that the modules cannot be resolved through modularity
optimization, not even if they were complete graphs connected by a
single link.  As we have seen from Eq.~(\ref{eq14}), it is possible to
miss modules of larger size, if they share more links with the rest of
the network (and with each other). For $l_1\neq l_2$ the conclusion is
similar but the scales $l_R^{min,max}$ are modified by simple factors.

\section{Consequences}
\label{sec5}

\begin{figure}[t]
\includegraphics[width=5.0cm]{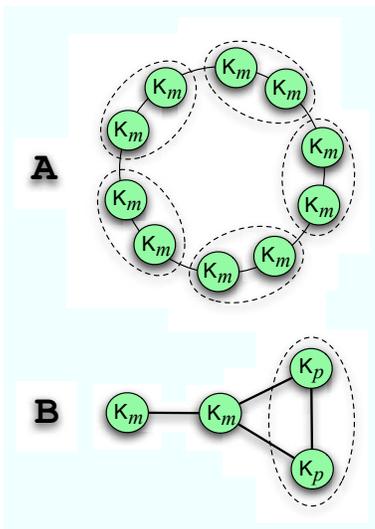}
\caption{\label{fig3} ({\bf A}) A network made out of identical
  cliques (which are here complete graphs with $m$ nodes) connected by
  single links.  If the number of cliques is larger than about
  $\sqrt{L}$, modularity optimization would lead to a partition where
  the cliques are combined into groups of two or more (represented by
  a dotted line).  ({\bf B}) A network with four pairwise identical cliques 
(complete graphs with $m$ and
  $p<m$ nodes, respectively); if $m$ is large enough with respect to $p$
  (e.g. $m=20$, $p=5$), modularity optimization merges the
  two smallest modules into one (shown with a dotted line).}
\end{figure}

We begin with a very schematic example, for illustrative purposes.  In
Fig.~\ref{fig3}(A) we show a network consisting of a ring of cliques,
connected through single links.  Each clique is a complete graph $K_m$
with $m$ nodes and has $m(m-1)/2$ links. If we assume that there are
$n$ cliques, with $n$ even, the network has a total of $N=nm$ nodes
and $L=nm(m-1)/2+n$ links.
The network has a clear modular structure where the communities
correspond to single cliques and we expect that any detection
algorithm should be able to detect these communities. The modularity
$Q_{single}$ of this natural partition can be easily calculated and
equals
\begin{equation}
Q_{single}=1-\frac{2}{m(m-1)+2}-\frac{1}{n}.
\label{eq17}
\end{equation}
On the other hand, the modularity $Q_{pairs}$ of the partition in
which pairs of consecutive cliques are considered as single
communities (as shown by the dotted lines in Fig.~\ref{fig3}(A)) is
\begin{equation}
Q_{pairs}=1-\frac{1}{m(m-1)+2}-\frac{2}{n}.
\label{eq16}
\end{equation}
The condition $Q_{single}>Q_{pairs}$ is satisfied if and only if 
\begin{equation}
m(m-1)+2>n.
\label{eq18}
\end{equation}
In this example, $m$ and $n$ are independent variables and we can
choose them such that the inequality of formula (\ref{eq18})
is not satistied. For instance, for $m=5$ and $n=30$, $Q_{single}=0.876$ and
$Q_{pairs}=0.888>Q_{single}$. An efficient algorithm looking for the maximum of
the modularity would find the configuration with pairs of cliques and
not the actual modules. The difference $Q_{pairs}-Q_{single}$ would be even larger
if $n$ increases, for $m$ fixed.

The example we considered was particularly simple and hardly
represents situations found in real networks. However, the initial
configuration that we considered in the previous section
(Fig.~\ref{fig2}) is absolutely general, and the results make us free
to design arbitrarily many networks with obvious community structures
in which modularity optimization does not recognize (some of) the real
modules.  Another example is shown in Fig.~\ref{fig3}(B). The circles
represent again cliques, i.e. complete graphs: the two on the left
have $m$ nodes each, the other two $p<m$ nodes.  If we take $m=20$ and
$p=5$, the maximal modularity of the network corresponds to the
partition in which the two smaller cliques are merged [as shown by the
dotted line in Fig.~\ref{fig3}(B)].
%
This trend of the optimal modularity to group small modules has
already been remarked in \cite{Muff:2005}, but as a result of
empirical studies on special networks, without any complete
explanation.  

In general, we cannot make any definite statement about modules
found through modularity optimization without a method which verifies
whether the modules are indeed single communities or a combination of
communities. It is then necessary to inspect the structure of each of
the modules found. As an example, we take the network of
Fig.~\ref{fig3}(A), with $n=30$ identical cliques, where each clique is
a $K_m$ with $m=5$.  As already said above, modularity optimization
would find modules which are pairs of connected cliques. By inspecting
each of the modules of the `first generation' (by optimizing
modularity, for example), we would ultimately find that each module is
actually a set of two cliques.

We thus have seen that modules identified through modularity
optimization may actually be combinations of smaller modules. During the process
of modularity optimization, it is favorable to merge connected 
modules if they are sufficiently small. 

We have seen in the previous Section that any two interconnected modules,
fuzzy or not, are merged if the number of links inside each of them
does not exceed $l_R^{min}$. This means that the largest 
structure one can form by merging a pair of modules of any type
(including cliques) has at least $2l_R^{min}$ internal links.
By reversing the argument, we conclude that 
if modularity optimization finds a module
$\cal S$ with $l_S$ internal links, it may be that the latter
is a combination of two or more smaller communities if
\begin{equation}
l_S< 2l_R^{min}=\sqrt{2L}.
\label{eq19}
\end{equation}
This example is an extreme case, in which the internal partition of
$\cal S$ can be arbitrary, as long as the pieces are modules in the
sense discussed in Section ~\ref{sec2}. Under the condition (\ref{eq19}), 
the module could in principle be a
cluster of loosely interconnected complete graphs.

On the other hand, the upper limit of $l_S$ can be much larger than
$\sqrt{2L}$, if the substructures are on average more interconnected
with each other, as we have seen in Section~\ref{sec4}.  In fact,
fuzzy modules can be combined with each other even if they contain
many more than $l_R^{min}$ links.  The more interconnected the
modules, the larger will be the resulting supermodule. In the extreme
case in which all submodules are very fuzzy, the size $l_S$ of the
supermodule could be in principle as large as that of the whole
network, i.e. $l_S<L$. This result comes from the extreme case where
the network is split in two very fuzzy communities, with $L/4$
internal links each and $L/2$ between them. By virtue of
Eq.~(\ref{eq14}), it is favorable (or just as good) to merge the two
modules and the resulting structure is the whole network. This limit
$l_S<L$ is of course always satisfied but suggests here that it is
important to carefully analyze all modules found through modularity
optimization, regardless of their size.

The probability that a very large module conceals substructures is
however small, because that could only happen if all hidden
submodules are very fuzzy communities, which is unlikely. Instead,
modules with a size $l_S \sim \sqrt{2L}$ or smaller can result from an
arbitrary merge of smaller structures, which may go from loosely
interconnected cliques to very fuzzy communities. Modularity
optimization is most likely to fail in these cases.

In order to illustrate this theoretical discussion, we analyze five
examples of real networks: 

\begin{enumerate}
\item{the transcriptional regulation network of
\textit{Saccharomyces cerevisiae} (yeast);}
\item{the transcriptional regulation network of \textit{Escherichia
  coli};}
\item{a network of electronic circuits;}
\item{a social network;}
\item{the neural network of \textit{Caenorhabditis Elegans}.}
\end{enumerate}

We downloaded the lists of edges of the first four networks from 
Uri Alon's
Website~\cite{alonwebsite}, whereas the last one was downloaded from the WebSite
of the Collective Dynamics Group at Columbia University~\cite{colweb}.

In the transcriptional regulation networks, 
nodes represent operons, i.e. groups of genes that are
transcribed on to the same mRNA and an edge is set between two nodes A
and B if A activates B. These systems have been
previously studied to identify motifs in complex networks~\cite{alon}.
There are $688$ nodes, $1,079$ links for
yeast, $423$ nodes and $519$ links for \textit{E. coli}.
Electronic circuits can be viewed as networks in which vertices are
electronic components (like capacitors, diodes, etc.) and connections are
wires. Our network maps one of the benchmark circuits 
of the so-called ISCAS'89 set; it has $512$ nodes, $819$ links.  
In the social network we considered, nodes are people of a group and links 
represent positive sentiments directed from one person to another, 
based on questionnaires: it has $67$ nodes and $182$ links.
Finally, the neural network of \textit{C. elegans} is made of $306$ nodes (neurons),
connected through $2,345$ links (synapsis, gap junctions). We remark that
most of these networks are directed, here we considered them as undirected.

First, we look for the modularity maximum by using simulated
annealing.  We adopt the same recipe introduced in
Ref.~\cite{Guimera:2005}, which makes the optimization procedure very
effective. There are two types of moves to pass from a network
partition to the next: individual moves, where a single node is passed
from a community to another, and collective moves, where a pair of
communities is merged into a single one or, vice versa, a community is
split into two parts. Each iteration at the same temperature consists
of a succession of $N^2$ individual and $N$ collective moves, where
$N$ is the total number of nodes of the network. The initial
temperature $T$ and the temperature reduction factor $f$ are
arbitrarily tuned to find the highest possible modularity: in most
cases we took $T\sim 1$ and $f$ between $0.9$ and $0.99$.

We found that all networks are characterized by high modularity
peaks, with $Q_{max}$ ranging from $0.4022$ (\textit{C. elegans}) to
$0.7519$ (\textit{E. coli}). The corresponding optimal partitions consist of
$9$ (yeast), $27$ (\textit{E. coli}), $11$ (electronic), $10$ (social) and 
$4$ (\textit{C. elegans}) modules (for \textit{E. coli} our
results differ but are not inconsistent with those obtained in~\cite{Guimera:2005} 
for a slighly different database; these differences
however do not affect our conclusions). In order to check if the
communities have a substructure, we used again modularity
optimization, by constraining it to each of the modules found. In all
cases, we found that most modules displayed themselves a clear community structure,
with very high values of $Q$.  The total number of submodules is 
$57$ (yeast), $76$ (\textit{E. coli}), $70$ (electronic), $21$ (social) and 
$15$ (\textit{C. elegans}), and is far larger
than the corresponding number at the modularity peaks.
The analysis of course is necessarily biased by the fact that we neglect
all links between the original communities, and it may be that the
submodules we found are not real modules for the original network. In
order to verify that, we need to check whether the condition of
Eq.~(\ref{eq2}) is satisfied or not for each submodule and we found that
it is the case. A further inspection of the communities found through
modularity optimization thus reveals that they are, in fact, clusters of
smaller modules.  The modularity values corresponding to the
partitions of the networks in the submodules are clearly smaller than the
peak modularities that we originally found
through simulated annealing (see Table~\ref{tab1}).
By restricting modularity optimization
to a module we have no guarantee that we accurately detect its substructure
and that this is a safe way to proceed.
Nevertheless, 
we have verified that all
substructures we detected are indeed modules, so our
results show that the search for the modularity optimum is not
equivalent to the detection of communities defined through
Eq.~(\ref{eq2}).  
\begin{table}[htbp]
\begin{center}
\begin{tabular}{|c||c|c|c|}
\hline
network    & $\#$ modules ($Q_{max}$) & total $\#$ of modules ($Q$) \\
\hline \hline 
Yeast   & 9 (0.7396)  & 57 (0.6770) \\
E. Coli & 27 (0.7519) & 76 (0.6615) \\
Electr. circuit   & 11 (0.6701)  & 70 (0.6401) \\
Social & 10 (0.6079) & 21 (0.5316) \\
C. elegans & 4 (0.4022) & 15 (0.3613) \\
\hline
\end{tabular}
\caption{\label{tab1} Results of the modularity analysis on real
  networks. In the second column, we report the number of modules
  detected in the partition obtained for the maximal modularity. These
  modules however contain submodules and in the third column we report
  the total number of submodules we found and the corresponding value of
  the modularity of the partition, which is {\it lower} than the peak 
  modularity initially found.}
\end{center}
\end{table}

The networks we have examined are fairly small but the problem we pointed
out can only get worse if we increase the network size, especially
when small communities coexist with large ones and the module size
distribution is broad, which happens in many
cases~\cite{Clauset:2004,Danon:2006}.  As an example, we take the
recommendation network of the online seller Amazon.com.  While buying
a product, Amazon recommends items which have been purchased by people
who bought the same product. In this way it is possible to build a
network in which the nodes are the items (books, music), and there is
an edge between two items $A$ and $B$ if $B$ was frequently purchased
by buyers of $A$.  Such a network was examined in
Ref.~\cite{Clauset:2004} and is very large, with $409,687$ nodes and
$2,464,630$ edges. The authors analyzed the community structure by
greedy modularity optimization which is not necessarily accurate but
represents the only strategy currently available for large
networks. They identified $1,684$ communities whose size distribution
is well approximated by a power law with exponent $2$. From the size
distribution, we estimated that over $95\%$ of the modules have sizes
below the limit of Eq.~(\ref{eq19}), which implies that basically all
modules need to be further investigated.

\section{Conclusions}
\label{sec6}

In this article we have analyzed in detail modularity and its
applicability to community detection. We have found that the
definition of community implied by modularity is actually not
consistent with its optimization which may favour network partitions
with groups of modules combined into larger communities. We could say that, 
by enforcing modularity optimization, the possible
partitions of the system are explored at a coarse level, so that
modules smaller than some scale may not be resolved. The resolution
limit of modularity does not rely on particular network structures,
but only on the comparison between the sizes of interconnected
communities and that of the whole network, where the sizes are
measured by the number of links.

The origin of the resolution scale lies in the fact that modularity is
a sum of terms, where each term corresponds to a module. Finding the
maximal modularity is then equivalent to look for the ideal tradeoff
between the number of terms in the sum, i.e. the number of modules,
and the value of each term. An increase of the number of modules does
not necessarily correspond to an increase in modularity because the
modules would be smaller and so would be each term of the sum. This is
why for some characteristic number of terms the modularity has a
peak. The problem is that this ``optimal'' partition, imposed by
mathematics, is not necessarily correlated with the actual community
structure of the network, where communities may be very heterogeneous
in size, especially if the network is large. 

Our result implies that modularity optimization might miss important
substructures of a network, as we have confirmed in real world
examples.  From our discussion we deduce that it is not possible to
exclude that modules of virtually any size may be clusters of modules,
although the problem is most likely to occur for modules with a number
of internal links of the order of $\sqrt{2L}$ or smaller.  For this
reason, it is crucial to check the structure of all detected modules,
for instance by constraining modularity optimization on each single
module, a procedure which is not safe but may give useful indications.

The fact that quality functions such as the modularity have an
intrinsic resolution limit calls for a new theoretical framework which
focuses on a local definition of community, regardless of its
size. Quality functions are still helpful, but their role should be
probably limited to the comparison of partitions with the same number
of modules.

\medskip Acknowledgments.\--- We thank A. Barrat, C. Castellano,
V. Colizza, A. Flammini, J. Kert\'esz and A. Vespignani for
enlightening discussions and suggestions. We also thank U. Alon for
providing the network data.




\end{document}